\documentclass[runningheads]{llncs}
\usepackage[T1]{fontenc}
\usepackage{subfig}
\usepackage{graphicx}
\usepackage{amsmath}
\usepackage{amssymb}
\usepackage{wrapfig}
\usepackage{verbatim}
\usepackage{upgreek}
\usepackage{booktabs}
\usepackage{multirow}
\usepackage{array}
\usepackage{listings}
\usepackage{fancyhdr}
\usepackage{tabularx}
\usepackage{comment}
\usepackage{booktabs} 
\usepackage{siunitx} 
\usepackage{hyperref}

\newcommand\asteriskfill{\leavevmode\xleaders\hbox{$\ast\ $}\hfill\kern0pt}

\begin{document}

\title{Spectral Graph Sample Weighting for Interpretable Sub-cohort Analysis in Predictive Models for Neuroimaging}
\titlerunning{Spectral Graph Sample Weighting for Neuroimaging}

\author{Magdalini Paschali\inst{1} \and
Yu Hang Jiang\inst{2} \and
Spencer Siegel\inst{2} \and
Camila Gonz\'{a}lez \inst{3} \and
\\Kilian M. Pohl\inst{3} \and
Akshay Chaudhari\inst{1,4} \and
Qingyu Zhao\inst{5}
}


%
\authorrunning{M. Paschali et al.}

\institute{Department of Radiology, Stanford University, Stanford, CA, USA
\\\href{mailto:paschali@stanford.edu}{paschali@stanford.edu}
\and
Department of Statistics, Stanford University, Stanford, USA
\and
Department of Psychiatry and Behavioral Sciences, \\Stanford University, Stanford, CA, USA
\and
Department of Biomedical Data Science, Stanford University, Stanford, CA, USA.
\and
Department of Radiology, Weill Cornell Medicine, New York, NY, USA
}

\maketitle

\begin{abstract} 
Recent advancements in medicine have confirmed that brain disorders often comprise multiple subtypes of mechanisms, developmental trajectories, or severity levels. Such heterogeneity is often associated with demographic aspects (e.g., sex) or disease-related contributors (e.g., genetics). Thus, the predictive power of machine learning models used for symptom prediction varies across subjects based on such factors. To model this heterogeneity, one can assign each training sample a factor-dependent weight, which modulates the subject's contribution to the overall objective loss function. 
To this end, we propose to model the subject weights as a linear combination of the eigenbases of a spectral population graph that captures the similarity of factors across subjects. In doing so, the learned weights smoothly vary across the graph, highlighting sub-cohorts with high and low predictability.  
Our proposed sample weighting scheme is evaluated on two tasks. First, we predict initiation of heavy alcohol drinking in young adulthood from imaging and neuropsychological measures from the National Consortium on Alcohol and NeuroDevelopment in Adolescence (NCANDA). Next, we detect Dementia \textit{vs.} Mild Cognitive Impairment (MCI) using imaging and demographic measurements in subjects from the Alzheimer’s Disease Neuroimaging Initiative (ADNI). Compared to existing sample weighting schemes, our sample weights improve interpretability and highlight sub-cohorts with distinct characteristics and varying model accuracy. 
\end{abstract}
\section{Introduction}
Sample weighting assigns different levels of importance across subjects to modulate their contribution towards training machine learning models~\cite{ren2018learning}. When used for predicting symptom outcomes from neuroimaging measures, these weights serve various purposes. For instance, they guide models to ignore outliers and noisy labels~\cite{Adeli2016,Roh2021}. Moreover, these weights can steer a model to focus on ``hard-to-classify'' samples~\cite{liu2021just} by deriving more accurate decision boundaries. Other uses include gradually increasing the complexity of training via curriculum learning~\cite{jiang2018mentornet,jiang2015self}, overcoming class imbalance and distribution shift~\cite{ren2018learning}, and achieving faster convergence~\cite{SANTIAGOLOW}. 

Recent studies have shown in multiple datasets~\cite{Greene2022,Dhamala2022} that when training classification models to predict phenotypic measures from neuroimaging data, the accuracy of models systematically varies with sociodemographic and clinical factors (e.g. race or level of education). To that end, in this paper, we rethink the utility and purpose of sample weighting in neuroimaging-based predictive modeling. Rather than reweighing challenging or underrepresented samples, we learn sample weights to gain population-level insights into the intrinsic relationship between predictive power and auxiliary factors characterizing brain disorders. The goal of our approach is to improve the interpretability of learned weights, highlighting specific sub-cohorts associated with more pronounced predictive cues.

\begin{figure}[!t]
\centering
\includegraphics[width=0.88\textwidth]{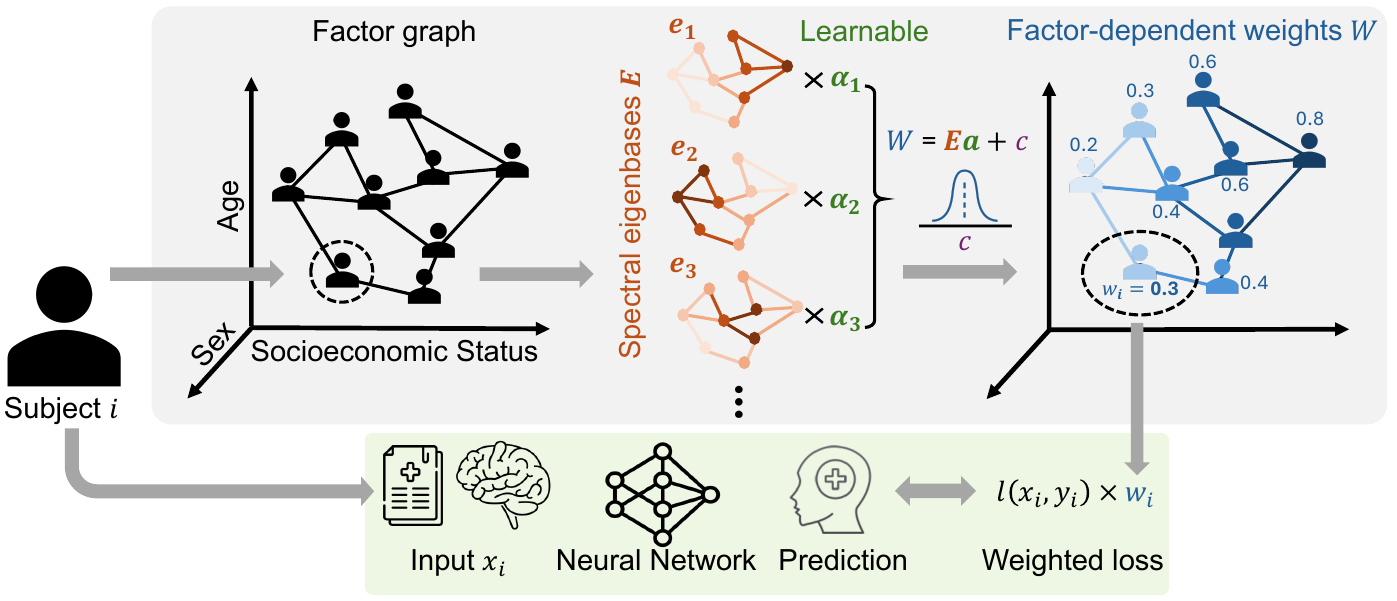}
\caption{We first create a population graph using pre-defined sociodemographic, genetic, or environmental factors that share associations with specific brain diseases. Next, a machine learning model is trained to predict symptom outcome given imaging and non-imaging data for each subject. During training, the classification loss is weighted by $w$, which is a linear combination of the graph eigenbases $\mathbf{E}$ with a learnable vector $\mathbf{a}$. The learned weights highlight sub-cohorts that share common characteristics and achieve higher predictive power.} 
\label{fig:overview}
\end{figure}

One existing solution to capturing factor-dependent predictive power is to learn a separate model in different sub-cohorts, e.g., using sex-specific models~\cite{Jiang2019,Dhamala2022HBM}. However, this approach becomes intractable as the number of sub-cohorts exponentially grows with the number of considered factors while the number of samples in each sub-cohort drastically decreases. Here, we propose a data-driven solution to reveal the interplay between model prediction and cohort-specific factors within a single coherent model. To do so, we explicitly associate sample weights with pre-defined selected sociodemographic, genetic, or environmental factors that share known associations with specific brain diseases. We consider a transductive setting where we first construct a factor graph that connects all samples based on the similarity of their factor values. As shown in Fig.~\ref{fig:overview}, the sample weights are enforced to be a linear combination of the spectral eigenbases of the graph Laplacian during the training of the classification model. Thus, the learned weights vary smoothly with respect to the selected factors. Through this process, our weights can identify meaningful sub-cohorts with distinct characteristics and varying classification accuracy scores.

We evaluate our approach on two public datasets. First, we use imaging and neuropsychological measures acquired in no-to-low drinking adolescents before they turn 18 years old from the National Consortium on Alcohol and NeuroDevelopment in Adolescence (NCANDA)~\cite{brown2015national}. We forecast which subjects will initiate heavy alcohol consumption after leaving high school and before the legal drinking age of 21 years in the United States~\cite{zhao2024identifying}. We consider sex, socioeconomic status (SES), and family alcohol history as factors associated with heavy drinking initiation~\cite{Collins2016,Tschorn2021}. Next, we use imaging and demographic measures from participants of the Alzheimer’s Disease Neuroimaging Initiative (ADNI) to predict Dementia \textit{vs.} Mild Cognitive Impairment (MCI). 
We select sex, age, and the apolipoprotein E gene $\epsilon$4 (APOE $\epsilon$4)
as factors associated with Alzheimer's Disease (AD), the most common type of dementia~\cite{saykin2010alzheimer,podcasy2016considering}. In both analyses, our approach learns meaningful weights informing which specific sub-cohorts are intrinsically associated with which predictive cues.

\section{Method}

In this section, we describe our sample-weighting approach. First, we construct a spectral factor graph on all study participants. Then, we train a neural network to predict symptom outcome and learn weights associated with the subjects based on the constructed population graph. We make a transductive assumption: while the prediction labels of testing samples are unknown, auxiliary factors of testing samples (e.g., age and sex) are known beforehand. This assumption is common in prospective studies of neurological or neuropsychiatric disorders, which aim to find population-level brain-phenotype relationships using machine learning~\cite{drysdale2017resting}, as opposed to deriving MRI-based diagnosis in hospital settings. Nevertheless, the factors are only used to determine sample weights and not as direct input to our machine learning model (no data leakage) as they can potentially confound the prediction. 

    \textbf{Parameterizing sample weights by factor graphs.} Let $\mathbf{X}=[x_1,...,x_N]^\top$ be the input data to a prediction model of $N$ independent samples (including both training and testing samples) and $\mathbf{y}=[y_1,...,y_N]$ be their corresponding target prediction labels. Our goal is to learn the weights $\mathbf{W} = [w_1, w_2, \dots, w_N]^\top$ to weigh the prediction loss $\sum l(x_i, y_i)$ (e.g., binary cross-entropy for a classification model) separately across training samples. 

Each sample is characterized by $D$ factors of interest $[s^1,...,s^D]$ that are associated with the predictive task of the model. To link the learned weights to the $D$ factors, we first construct a factor graph represented by a connectivity matrix $A\in \mathbb{R}^{N\times N}$, where $A_{i,j}$ encodes the similarity between sample $i$ and $j$ in terms of their factor values. We normalize each $s^d$ to z-scores and construct 
\begin{equation}
    A_{i,j}= 
\begin{cases}
    \frac{1}{\sum_d (s^d_i - s^d_j)^2 + 1}& \text{if } i \in \mathcal{N}(j) \text{ or } j\in \mathcal{N}(i) \\
    0              & \text{otherwise,}
\end{cases}
\end{equation}
where $\mathcal{N}(i)$ defines the set of $K$ nearest neighbors of sample $i$. Based on spectral graph theory~\cite{Chung1997}, we derive the first $M$ eigenbases with non-zero eigenvalues from the Laplacian matrix of $A$.
\begin{equation}
    \mathbf{E} = \begin{bmatrix}
        e_{11} & \dots & e_{1M}\\
        \vdots & \ddots & \vdots\\
        e_{N1} & \dots & e_{NM}
    \end{bmatrix} \in \mathbb{R}^{N\times M}.
\end{equation} 
Here, all eigenbases (columns of $\mathbf{E}$) are mutually orthogonal. Each eigenbasis has a zero-sum $\sum_i e_{ij}=0$ and encodes a major mode of variation of the graph (i.e., low-frequency Fourier bases defined on the graph). As such, we can re-parameterize the sample weights $\mathbf{W}$ via a learnable vector \(\mathbf{a} = [a_1, a_2, \dots, a_M]\) as
$
\mathbf{W} = c + \mathbf{Ea},
$
where \(c\) is a constant representing the centering reference weight. In doing so, the sample weights are enforced to smoothly vary across the graph by only containing low-frequency variation. 

\textbf{Transductive learning of sample weights.} With the above reparameterization, we can compute $\mathbf{W}$ of all samples through optimizing the learnable vector $\mathbf{a}$ only on the training set.
 
Without loss of generality, we assume that of the $N$ samples, the first $N'$ belong to the training set and the remaining $N''$ samples to the testing set. Then the eigenbases and weights can be written as a vertical concatenation of training and testing samples, $\mathbf{E}=[\mathbf{E}';\mathbf{E}''], \mathbf{W}=[\mathbf{W}';\mathbf{W}'']$. We now use $\mathbf{E}'$ to learn the weights associated with the $N'$ training samples by minimizing their weighted prediction loss
\begin{equation}
L(\mathbf{X}', \mathbf{y}', \mathbf{a}) = \sum_i^{N'} w_i l(x_i, y_i) + \sum_i^{N'} \text{max}(0,-w_i)\text{, s.t., }\mathbf{W'} = c + \mathbf{E'a},
\end{equation}
where the second summation penalizes negative sample weights. After training, we apply the learned $\mathbf{a}$ to infer the weights of testing samples $\mathbf{W''} = c + \mathbf{E''a}$. With this setting, all imaging and non-graph-related measurements of the testing samples remain unseen during training and do not influence the optimization of the neural network and $\mathbf{a}$.

\section{Experimental Settings}

\noindent\textbf{Datasets.} The NCANDA study~\cite{brown2015national} recruited 831 youths across five sites in the United States and performed annual imaging, behavioral, and neuropsychological assessments~\cite{brown2015national}. According to the youth-adjusted Cahalan score \cite{Pfefferbaum2017}, 207 remained no-to-low drinking from their baseline visits to age 21 years and 192 participants were no-to-low drinkers before age 18 years but initiated heavy or binge drinking behaviors before the legal drinking age of 21 years. Our goal is to predict which of the 399 participants (206 female/193 male) will initiate heavy drinking between age 18 and 21 years based on their longitudinal assessments before age 18 years (age at baseline: 12 - 18 years,  2.82 $\pm$ 1.47 assessments per participant). We use $172$ measurements at each assessment, of which $145$ are non-imaging scores capturing demographic, life experiences, personality, neuropsychological, and behavioral measures~\cite{Paschali2022}. Additionally, the average fractional anisotropy (FA) of $28$ brain regions defined by the JHU Atlas are derived from each Diffusion Tensor Image after being processed by the publicly available NCANDA diffusion pipeline~\cite{Pohl2016}. To construct the factor graph in this study, we consider sex, SES (as defined by parental years of education), and family alcohol history (which accounts for the number of first- and second-degree relatives with alcohol use disorder).

Additionally, we evaluate our method on 1191 subjects from ADNI 1,2,3,GO \cite{petersen2010alzheimer} (503 female/688 male) with the goal to predict Dementia (N=507) \textit{vs.} MCI (N=684). We utilize longitudinal assessments from each subject (age at baseline: 55 - 95 years, 2.85 $\pm$ 1.79 visits per subject). For each assessment, we use 314 T1-weighted MRI Freesurfer scores from the UCSF  release~\cite{hartig2014ucsf} and information regarding their education, ethnicity, race, and marital status. We only utilize samples whose imaging measurements passed all quality assessments to avoid noisy or erroneous information. For the graph construction, we select the risk factors of sex, age, and the possession of the APOE $\epsilon$4 gene.

\noindent\textbf{Implementation.}
To model the longitudinal nature of the datasets, we use a deep learning model combining a Gated Recurrent Unit~\cite{cho2014learning} layer and two Fully Connected layers. Our model uses all available yearly subject assessments and generates predictions in a sequence-to-one fashion. For NCANDA, the models are trained for 100 epochs with an initial learning rate of 1e-4. For ADNI we train for 30 epochs with an initial learning rate of 1e-3. Both models use the Adam Optimizer; the learning rate for the trainable weight vector $\mathbf{a}$ is 1e-5. All models are implemented in PyTorch 1.13.1, and the code is publicly available~\footnote{\url{https://github.com/MaggiePas/sample_weighting}}. 

The average and standard deviation of balanced accuracy (BACC) and F1-score are reported for 5-fold subject-level stratified cross-validation across all experiments. Based on identifying the significant change point in the eigenvalues associated with the eigenbases~\cite{james2013introduction}, we set $M=13$ eigenbases for NCANDA and $M=7$ for ADNI. To identify the model with the most distinct and separable learned-weight sub-cohorts, we measure the absolute difference of BACC in percentage between the subjects with high \textit{vs.} low weights. We evaluate the impact of the number of neighbors $K$ used to construct the factor graph by training models with 10, 30, 50, 75, and 100 neighbors. We also evaluate the impact of 5 different options for the centering hyperparameter $c$ combined with each neighbor choice.

\noindent\textbf{Baselines}. To the best of our knowledge, there are no directly comparable baselines that learn sample weights related to auxiliary factors. In addition, existing methods cannot generate sample weights for testing samples. We compare the proposed sample weighting approach with a baseline model trained without any weighting and with two baseline methods that learn sample weights during training without considering auxiliary factors. Just Train Twice (JTT)~\cite{liu2021just} first trains a classification model and, in a second step, trains a second one that up-weights the training samples misclassified by the first model. Thus, the sample weight is equal to 1 if a training sample is correctly classified by the first model and equal to $\lambda$ if it is misclassified, where $\lambda$ represents the occurrence of the up-weighted sample in the next training cycle. For both datasets, we set $\lambda=2$ since it achieves the highest BACC. We also compare our method against a sample re-weighting meta-learning approach that learns to assign weights to training examples based on their gradient directions~\cite{ren2018learning}. Finally, we compare our approach with one variation where the trainable vector $\mathbf{a}$ is kept constant to 1s during training. This experiment (Only Graph in Table~\ref{tab:my_label}) showcases the difference in BACC when the sample weights are derived only from the factor graph without a learnable component that is updated during training.

\begin{table}[t]
    \caption{Balanced accuracy (BACC) and F1-score across 5-fold stratified cross-validation for the proposed model and baseline approaches.}
    \centering
\resizebox{0.84\textwidth}{!}{
    \begin{tabular}{w{l}{2.2cm}w{c}{2cm}w{c}{2cm}w{c}{2cm}w{c}{2cm}}
         &  \multicolumn{2}{c}{\textbf{NCANDA}}&  \multicolumn{2}{c}{\textbf{ADNI}}\\
 & \textbf{BACC}& \textbf{F1} & \textbf{BACC} & \textbf{F1} 
 \\ \hline
         No weigthing &  61.4 $\pm$ 3.6 & 61.3 $\pm$ 3.6 & 62.1 $\pm$ 9.50 & 61.1 $\pm$ 12.4\\
         JTT & 62.1 $\pm$ 3.0 &  60.6 $\pm$ 3.7 &  68.9 $\pm$ 8.3 &  67.7 $\pm$ 9.6\\
         Meta-weighting &  61.0 $\pm$ 5.0&  60.1 $\pm$ 5.4&  65.2 $\pm$ 10.0&  64.6 $\pm$ 12.3\\
         Only Graph &  62.2 $\pm$ 2.8 &  62.0 $\pm$ 2.8&  68.4 $\pm$ 7.1 &  67.4 $\pm$ 7.5\\
         \textbf{Ours} &  63.7 $\pm$ 3.7 &  63.5 $\pm$ 3.6&  68.3 $\pm$ 7.0 &  67.3 $\pm$ 7.3\\ \hline
    \end{tabular}
    }
    \label{tab:my_label}
\end{table}

\section{Results and Discussion}

\begin{figure}[t]
\centering
  \includegraphics[width=0.87\linewidth]{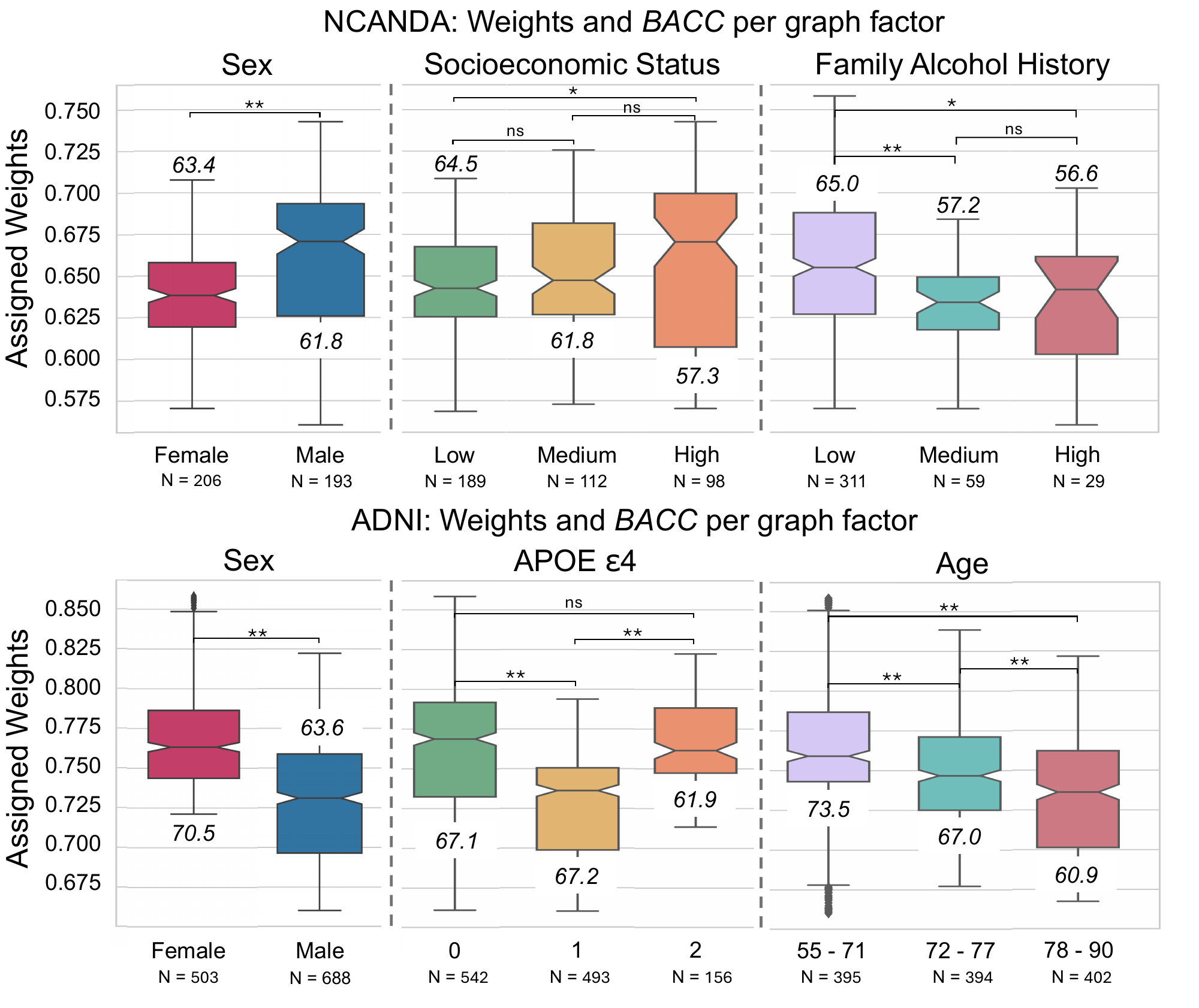}
  \caption{Comparison of learned weights across cohorts by graph factors for NCANDA and ADNI. The statistical difference of the weights across cohorts is measured with the Mann-Whitney U-test. **:p<0.001, *:p<0.05, ns:p>0.05. The \textit{BACC} for each sub-cohort is shown under or above the box.}
  \label{fig:figure2}
\end{figure}

\textbf{Overall accuracy comparison.} Table~\ref{tab:my_label} summarizes the accuracy scores for each method and dataset. compared to the model without sample weighting, the proposed sample weighting positively impacted the overall BACC, achieving a 3.8\% (2.3 BACC points) increase for NCANDA and a 10\% improvement in ADNI (6.2 BACC points). This showcases that sample weighting using factors associated with the target of the ML model improves predictive power. Next, we compare our model with JTT. For NCANDA, our model achieves 2.1\% higher BACC, and for ADNI a marginally 0.8\% lower BACC. JTT assigns high weights to hard-to-classify samples. This can improve overall BACC, but lacks interpretability of the learned weights. Meta-weighting is outperformed by our graph-weighting approach by 4.4\% for NCANDA and 4.7\% for ADNI in terms of BACC. This highlights the benefits of learned weights associated with target-specific risk factors compared to weights that aim to lower the overall loss. Finally, when we compare the two variations of our method with and without trainable $\mathbf{a}$ vector, we notice an improvement of 2.4\% for NCANDA while for ADNI the models perform almost identically (68.3 \textit{vs.} 68.4 BACC). 

\noindent\textbf{Sub-cohort analysis.} Our method is capable of associating sample weights with predictability across sub-cohorts. Comparing the BACC on the sub-cohorts with weights over the median weight (high) and weights under the median (low), the proposed model achieved 66.5 on high- and 60.5 BACC on low-weight cohorts for NCANDA and 72.3 \textit{vs.} 64.9 for ADNI. 
This indicates a substantial difference in the model's predictive power between the two sub-cohorts, distinguished by high- vs. low-weights.

Furthermore, using the graph factors, we can further evaluate model accuracy and weight variations across fine-grained groups, shown in Fig.~\ref{fig:figure2}. First, in both datasets, we observed significant differences in learned weights between male and female subjects (Mann-Whitney U-test, p<0.001). Notably, females consistently demonstrated higher BACC than males across both datasets.

With respect to SES in NCANDA, the model weights increase along with the SES of the subjects, and the accuracy is the highest for subjects with low SES. Lastly, sample weights decrease as family alcohol history increases in NCANDA, with the model achieving the highest BACC for subjects with low family history. These findings comport with the literature that SES and family history are key risk factors for developing drinking problems in both adolescents and adults~\cite{Collins2016,Tschorn2021} and highlight the need for designing sex-specific preventative programs during adolescence to lower the impact of alcohol misuse in young adulthood~\cite{Dir2017}.

In ADNI, the subject weights decrease significantly (p<0.001) as age increases from 55-71 years to 78-90 years. Along with the sample weights, the model BACC also decreases from 73.5 to 60.9 across age sub-cohorts. The age-dependent predictive power aligns with existing evidence that early-onset dementia in younger individuals has been associated with more widespread cortical atrophies, with their effect being more prominent than older individuals~\cite{mendez2019early}. Moreover,
subject weights are significantly different (p<0.001) for subjects with 0 \textit{vs.} 1 and 1 \textit{vs.} 2 counts of $\epsilon4$ alleles (APOE $\epsilon4$). This gene is a known genetic risk factor for AD with a two- to three-fold increased risk in subjects with 1 $\epsilon4$ allele that rises to 12-fold in those with 2 alleles~\cite{saykin2010alzheimer}. Our model achieves 8.5\% higher BACC for subjects with 0 or 1 copy of the APOE $\epsilon4$ gene than those with 2 copies of the gene. 
Interestingly, research suggests that possessing one or two copies of APOE $\epsilon4$ can advance the onset age of dementia by 5 to 10 years~\cite{belloy2020association}, potentially impacting model predictions. 
Overall, our approach manages to identify meaningful sub-cohorts with high predictability across datasets. 

\begin{figure}[t]
\centering
  \includegraphics[width=0.84\linewidth]{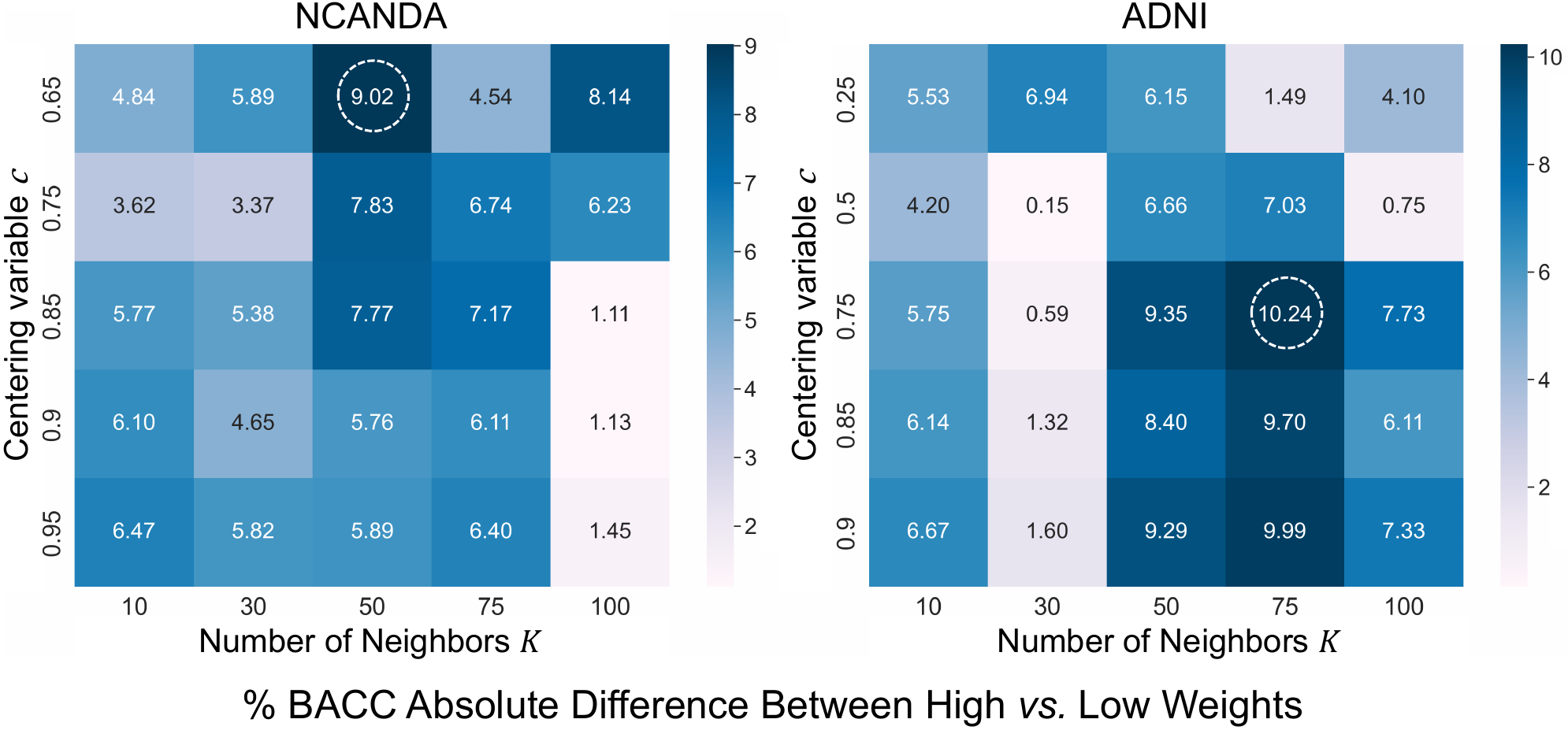}
  \caption{Impact of choice of neighbors and centering hyperparameter for NCANDA and ADNI. We measure the absolute difference in \% of BACC between the cohorts with high \textit{vs.} low weights to compare the ability of different models to create distinct and highly separable sub-cohorts based on the learned weights.}
  \label{fig:neighbors_centering}
\end{figure}
\noindent\textbf{Impact of neighbors and centering.} In Fig.~\ref{fig:neighbors_centering} we show the absolute difference (\%) in BACC between cohorts with high \textit{vs.} low weights. For both datasets, using $K=50$ and $K=75$ neighbors for the factor graph produced the highest gap between the sub-cohorts' BACC, regardless of the hyperparameter $c$. Regarding centering, models with $c$ larger than 0.5 achieved the highest gaps, with values of $c$ ranging from 0.65 to 0.75 being the most advantageous. Specifically for NCANDA, 50 neighbors and $c$=0.65 achieved a 9.02\% gap between high- and low-weight cohorts, while for ADNI, 75 neighbors and $c$=0.75  reached a 10.24\% gap. Overall, these two hyperparameters played a role in the separability of cohort by weight but varied smoothly across combinations. Specifically, all models demonstrated consistent overall BACC performance, with ranges of 66\% to 69\% for ADNI and 62\% to 64\% for NCANDA, across various values of $K$ and $c$.

\section{Conclusion}
In this work, we proposed a method for learning sample weights linked to factors related to disease prediction using imaging and neuropsychological scores.
By parameterizing the weights as a linear combination of the eigenbases of a spectral population graph, our method learnt separable sample weights, improved interpretability, and identified sub-cohorts that were more informative for the classification models. We demonstrated the effectiveness of our approach on two tasks, achieving higher accuracy over the baseline without any weighting and highlighting sub-cohorts with distinct characteristics and predictive power. Future work could explore using the adjacency matrix derived from more features determined through feature importance analysis and literature. Moreover, our approach could be validated on datasets beyond neuroimaging and explore further model architectures.

\section*{Acknowledgements}
This work was supported by the U.S. National Institute (DA057567, AA021697, AA010723, AA028840), BBRF Young Investigator Grant, the DGIST Joint Research Project, the 2024 HAI Hoffman-Yee Grant, and the HAI-Google Cloud Credits Award. The NCANDA data were based on a formal, locked data release NCANDA\_NIAAADA\_BASE\_V01,  NCANDA\_NIAAADA\_01Y\_V01 to \\NCANDA\_NIAAADA\_07Y\_V01 available via \url{https://nda.nih.gov/edit_collection.html?id=4513}. NCANDA data collection and distribution were supported by NIH funding AA021681, AA021690, AA021691, AA021692, {AA021\-695}, AA021696, AA021697.

%

\bibliographystyle{splncs04}
\bibliography{miccai.bib}

\begin{thebibliography}{10}
\providecommand{\url}[1]{\texttt{#1}}
\providecommand{\urlprefix}{URL }
\providecommand{\doi}[1]{https://doi.org/#1}

\bibitem{Adeli2016}
Adeli, E., Shi, F., An, L., Wee, C.Y., Wu, G., Wang, T.: Joint feature-sample selection and robust diagnosis of parkinson's disease from {MRI} data. NeuroImage  \textbf{141} (06 2016)

\bibitem{belloy2020association}
Belloy, M.E., Napolioni, V., Han, S.S., Le~Guen, Y., Greicius, M.D., Initiative, A.D.N., et~al.: Association of {Klotho-VS heterozygosity with risk of Alzheimer} disease in individuals who carry {APOE4}. JAMA neurology  \textbf{77}(7),  849--862 (2020)

\bibitem{brown2015national}
Brown, S.A., Brumback, T., Tomlinson, K., Cummins, K., Thompson, W.K., Nagel, B.J., De~Bellis, M.D., Hooper, S.R., Clark, D.B., Chung, T., et~al.: The {N}ational {C}onsortium on {A}lcohol and {N}eurodevelopment in {A}dolescence ({NCANDA}): a multisite study of adolescent development and substance use. Journal of studies on alcohol and drugs  \textbf{76}(6),  895--908 (2015)

\bibitem{cho2014learning}
Cho, K., Van~Merri{\"e}nboer, B., Gulcehre, C., Bahdanau, D., Bougares, F., Schwenk, H., Bengio, Y.: Learning phrase representations using {RNN} encoder-decoder for statistical machine translation. arXiv preprint arXiv:1406.1078  (2014)

\bibitem{Chung1997}
Chung, F.: Spectral Graph Theory. American Mathematical Society, Providence, R.I. (1997)

\bibitem{Collins2016}
Collins, S.: Associations between socioeconomic factors and alcohol outcomes. Alcohol research: current reviews  \textbf{38},  83--94 (04 2016)

\bibitem{Dhamala2022HBM}
Dhamala, E., Jamison, K.W., Jaywant, A., Kuceyeski, A.: Shared functional connections within and between cortical networks predict cognitive abilities in adult males and females. Human brain mapping  \textbf{43},  1087--1102 (2 2022), \url{https://pubmed.ncbi.nlm.nih.gov/34811849/}

\bibitem{Dhamala2022}
Dhamala, E., Yeo, B., Holmes, A.: One size does not fit all: Methodological considerations for brain-based predictive modeling in psychiatry. Biological Psychiatry  \textbf{93} (09 2022)

\bibitem{Dir2017}
Dir, A., Bell, R., Adams, Z., Hulvershorn, L.: Gender differences in risk factors for adolescent binge drinking and implications for intervention and prevention. Frontiers in Psychiatry  \textbf{8} (12 2017)

\bibitem{drysdale2017resting}
Drysdale, A.T., Grosenick, L., Downar, J., Dunlop, K., Mansouri, F., Meng, Y., Fetcho, R.N., Zebley, B., Oathes, D.J., Etkin, A., et~al.: Resting-state connectivity biomarkers define neurophysiological subtypes of depression. Nature medicine  \textbf{23}(1),  28--38 (2017)

\bibitem{Greene2022}
Greene, A., Shen, X., Noble, S., Horien, C., Hahn, C.A., Arora, J., Tokoglu, F., Spann, M., Carrion, C., Barron, D., Sanacora, G., Srihari, V., Woods, S., Scheinost, D., Constable, R.: Brain–phenotype models fail for individuals who defy sample stereotypes. Nature  \textbf{609},  1--10 (08 2022)

\bibitem{hartig2014ucsf}
Hartig, M., Truran-Sacrey, D., Raptentsetsang, S., Simonson, A., Mezher, A., Schuff, N., Weiner, M., et~al.: {UCSF} freesurfer methods. ADNI Alzheimers Disease Neuroimaging Initiative: San Francisco, CA, USA  (2014)

\bibitem{james2013introduction}
James, G., Witten, D., Hastie, T., Tibshirani, R., et~al.: An introduction to statistical learning, vol.~112. Springer (2013)

\bibitem{jiang2015self}
Jiang, L., Meng, D., Zhao, Q., Shan, S., Hauptmann, A.: Self-paced curriculum learning. In: Proceedings of the AAAI Conference on Artificial Intelligence. vol.~29 (2015)

\bibitem{jiang2018mentornet}
Jiang, L., Zhou, Z., Leung, T., Li, L.J., Fei-Fei, L.: Mentornet: Learning data-driven curriculum for very deep neural networks on corrupted labels. In: International conference on machine learning. pp. 2304--2313. PMLR (2018)

\bibitem{Jiang2019}
Jiang, R., Calhoun, V., Fan, L., Zuo, N., Jung, R., Qi, S., Lin, D., Li, J., Zhuo, C., Song, M., Fu, Z., Jiang, T., Sui, J.: Gender differences in connectome-based predictions of individualized intelligence quotient and sub-domain scores. Cerebral Cortex  \textbf{30} (07 2019)

\bibitem{liu2021just}
Liu, E.Z., Haghgoo, B., Chen, A.S., Raghunathan, A., Koh, P.W., Sagawa, S., Liang, P., Finn, C.: Just train twice: Improving group robustness without training group information. In: International Conference on Machine Learning. pp. 6781--6792. {PMLR} (2021)

\bibitem{mendez2019early}
Mendez, M.F.: Early-onset {Alzheimer} disease and its variants. Continuum (Minneapolis, Minn.)  \textbf{25}(1), ~34 (2019)

\bibitem{Paschali2022}
Paschali, M., Kiss, O., Zhao, Q., Adeli, E., Podhajsky, S., Müller-Oehring, E., Gotlib, I., Pohl, K., Baker, F.: Detecting negative valence symptoms in adolescents based on longitudinal self-reports and behavioral assessments. Journal of Affective Disorders  \textbf{312},  30--38 (2022)

\bibitem{petersen2010alzheimer}
Petersen, R.C., Aisen, P.S., Beckett, L.A., Donohue, M.C., Gamst, A.C., Harvey, D.J., Jack, C.R., Jagust, W.J., Shaw, L.M., Toga, A.W., et~al.: Alzheimer's disease neuroimaging initiative ({ADNI}): clinical characterization. Neurology  \textbf{74}(3),  201--209 (2010)

\bibitem{Pfefferbaum2017}
Pfefferbaum, A., Kwon, D., Brumback, T., Thompson, W., Cummins, K., Tapert, S., Brown, S., Colrain, I., Baker, F., Prouty, D., de~Bellis, M., Clark, D., Nagel, B., Chu, W., Park, S., Pohl, K., Sullivan, E.: Altered brain developmental trajectories in adolescents after initiating drinking. The American journal of psychiatry  \textbf{175}(4),  370--380 (2017)

\bibitem{podcasy2016considering}
Podcasy, J.L., Epperson, C.N.: Considering sex and gender in {Alzheimer} disease and other dementias. Dialogues in clinical neuroscience  \textbf{18}(4),  437--446 (2016)

\bibitem{Pohl2016}
Pohl, K., Sullivan, E., Rohlfing, T., Chu, W., Kwon, D., Nichols, B., Zhang, Y., Brown, S., Tapert, S., Cummins, K., Thompson, W., Brumback, T., Colrain, I., Baker, F., Prouty, D., de~Bellis, M., Voyvodic, J., Clark, D., Schirda, C., Pfefferbaum, A.: Harmonizing {DTI} measurements across scanners to examine the development of white matter microstructure in 803 adolescents of the {NCANDA} study. NeuroImage  \textbf{130},  194--213 (2016)

\bibitem{ren2018learning}
Ren, M., Zeng, W., Yang, B., Urtasun, R.: Learning to reweight examples for robust deep learning. In: International conference on machine learning. pp. 4334--4343. PMLR (2018)

\bibitem{Roh2021}
Roh, Y., Lee, K., Whang, S.E., Suh, C.: Sample selection for fair and robust training. In: Neural Information Processing Systems (NeurIPS) (2021)

\bibitem{SANTIAGOLOW}
Santiago, C., Barata, C., Sasdelli, M., Carneiro, G., Nascimento, J.C.: {LOW}: Training deep neural networks by learning optimal sample weights. Pattern Recognition  \textbf{110},  107585 (2021)

\bibitem{saykin2010alzheimer}
Saykin, A.J., Shen, L., Foroud, T.M., Potkin, S.G., Swaminathan, S., Kim, S., Risacher, S.L., Nho, K., Huentelman, M.J., Craig, D.W., et~al.: Alzheimer's disease neuroimaging initiative biomarkers as quantitative phenotypes: Genetics core aims, progress, and plans. Alzheimer's \& Dementia  \textbf{6}(3),  265--273 (2010)

\bibitem{Tschorn2021}
Tschorn, M., Lorenz, R., O'Reilly, P., Reichenberg, A., Banaschewski, T., Bokde, A., Quinlan, E., Desrivières, S., Flor, H., Grigis, A., Garavan, H., Gowland, P., Ittermann, B., Martinot, J.L., Artiges, E., Nees, F., Papadopoulos~Orfanos, D., Poustka, L., Millenet~(geb. Steiner), S., Rapp, M.: Differential predictors for alcohol use in adolescents as a function of familial risk. Translational Psychiatry  \textbf{11} (06 2021)

\bibitem{zhao2024identifying}
Zhao, Q., Paschali, M., Dehoney, J., Baker, F.C., de~Zambotti, M., De~Bellis, M.D., Goldston, D.B., Nooner, K.B., Clark, D.B., Luna, B., et~al.: Identifying high school risk factors that forecast heavy drinking onset in understudied young adults. Developmental Cognitive Neuroscience p. 101413 (2024)

\end{thebibliography}
\end{document}